# A phase field approach to no slip boundary conditions in dissipative particle dynamics and other particle models for fluid flow in geometrically complex confined systems

Zhijie Xu[1,a)] and Paul Meakin[2,3,4],

1. Energy Resource Recovery & Management, Idaho National Laboratory, Idaho Falls, Idaho 83415 USA

2. Center for Advanced Modeling and Simulation, Idaho National Laboratory, Idaho Falls, Idaho 83415 USA

3. Physics of Geological Processes, University of Oslo, Oslo 0316, Norway

4. Multiphase Flow Assurance Innovation Center, Institute for Energy Technology, Kjeller 2027, Norway

Dissipative particle dynamics (DPD) is an effective mesoscopic particle model with a lower computational cost than molecular dynamics because of the soft potentials that it employs. However, the soft potential is not strong enough to prevent the DPD particles that are used to represent the fluid from penetrating solid boundaries represented by stationary DPD particles. A phase field variable, $\phi(\mathbf{x},t)$, is used to indicate the phase at point **x** and time *t*, with a smooth transition from -1 (phase 1) to +1 (phase 2) across the interface. We describe an efficient implementation of no-slip boundary conditions in DPD models that combines solid-liquid particle-particle interactions with reflection at a sharp boundary located with subgrid scale accuracy using the phase field. This approach can be used for arbitrarily complex flow geometries and other similar particle models (such as smoothed particle hydrodynamics), and the validity of the model is demonstrated by DPD simulations of flow in confined systems with various geometries.

---

a) Electronic mail: zhijie.xu@inl.gov

# I. INTRODUCTION

The physics of complex fluids, multiphase fluids and the non-hydrodynamic behavior of fluids and fluid mixtures is an area of great current interest and practical importance. Examples include polymer solutions,[1, 2] colloidal suspensions,[3, 4] phase separation,[5] wetting phenomena,[6] multiphase fluids in fractured and porous media,[7] emulsions and microemulsions,[8] and small scale fluid dynamics where thermal fluctuation play an important role.[9] Computers simulations based on molecular dynamics, Monte Carlo methods, phase field models, lattice Boltzmann models and other approaches have played a key role in the development of a better understanding of the behavior of these complex fluid systems. Dissipative particle dynamics (DPD) is another promising approach to this class of problems.

DPD, a stochastic Lagrangian approach introduce by Hoogerbrugge and Koelman in 1992,[10] is based on the idea that particles can be used to represent clusters of atoms or molecules instead of single atoms or molecule to provide a simple and robust way of coarse graining the molecular dynamics of dense liquid systems. Because of the internal degrees of freedom associated with individual DPD particles, the DPD particle-particle interactions include dissipative and fluctuating interactions (related by the fluctuation-dissipation theorem),[11, 12] in addition to the conservative particle-particle interactions, and these interactions function as a thermostat for the model. The grouping of atoms or molecules into a single DPD particle (coarse graining) leads to averaged effective conservative interaction potentials (soft repulsive-only potentials in the standard DPD model) between the DPD particles. Consequently, the computational cost is substantially lowered due to the soft potentials as well as the coarse graining, and the computational advantage of DPD over molecular dynamics (MD) is about $1000\, N_m^{5/3}$, where $N_m$ is the number of atoms coarse-grained into a single DPD particle.[13] DPD is an effective mesoscale particle simulation technique for complex fluids on length and time scales, that are large compared with those accessible to fully atomistic MD simulations. However, for DPD simulations with interaction parameters that have been selected so that the DPD fluid properties match the properties of liquids such as water under standard temperature and pressure conditions, the degree of coarse graining, $N_m$, that can be used without forcing the system through a Kirkwood Alder transition[14, 15] to a solid state (because the coarse graining increases the magnitude of the conservative interactions energies relative to the thermal energy) or generating other artifacts is quite limited ( $N_m \approx 10$ ).[13, 16] Consequently, computational speedups (relative to molecular dynamics) greater than about $10^5$ cannot be achieved using the standard DPD model. Limited coarse graining such as replacing $CH_2$ by a single particle has been used for many years in molecular dynamics simulations, as has the use of thermostats for nonequilibrium molecular dynamics simulations.

DPD has been extensively used to simulate the bulk properties of complex fluids using periodic boundary conditions. However, in many important applications, such as the flow of suspensions and solutions of macromolecules though micro channels, and multiphase fluid flow through fractures, fracture junctions and porous media, where solid boundaries play a critical role, implementation of appropriate boundary conditions at the liquid-solid interface becomes an important issue in DPD simulations. One attractive approach is to use stationary DPD particles to represent the effects of confining solids on the fluid(s). This can be easily implemented, it allows complex wetting behaviors to be simulated, and it is closely related to the molecular interactions between fluids and solids. However, the soft interactions between the DPD particles that represent solid and liquid phases are not sufficient to prevent the fluid

DPD particles from penetrating through the fluid-solid interface into the solid. A variety of DPD boundary condition models have been described in the literature,[17-20] and they can be classified into the following four categories,

1. Periodic and Lees-Edwards (periodic shear strain) boundary conditions, which do not require explicit modeling of the solid boundary.
2. Use of "frozen" DPD particles inside the solid region to mimic solid walls.
3. Application of various collision algorithms at sharp boundaries, such as specular reflection, bounce-back, and Maxwellian reflection.
4. A combination of 2 and 3.

The Lees-Edward boundary method is used to simulate the effects of planar shear strain on complex fluids. It uses modified periodic boundary conditions to simulate the shear flow without explicitly modeling of the solid boundary. However, it cannot be applied to flow in confined systems or flow between moving walls with non-planar geometries. The use of frozen DPD particles to represent solid region is a popular way to create solid boundaries, and this approach has been used earlier in molecular dynamics simulations. The frozen particles can be placed either on a regular lattice or off lattice by simulating a dense fluid and freezing the particles after the system has equilibrated. Collision with sharp boundaries is easily implemented for simple geometries, and it has the advantage that the position of the boundary is well defined. Under equilibrium conditions, the collisions between DPD particles and the boundaries must be consistent with the principal of microscopic reversibility and the second law of thermodynamics.[21-23] The combination of fluid-solid particle-particle interactions and reflection at sharp boundaries allows the effects of physical and chemical heterogeneity on complex wetting behaviors to be investigated and captured in computer simulations while the loss of fluid particles due to penetration into the solid is prevented. Revenga, *et al*. [20] have discussed and compared various reflection mechanisms. In specular reflection, the velocity component tangential to the interface does not change but the normal component is reversed, while all velocity components are reversed for bounce-back reflection. In Maxwellian reflection, DPD particles are reintroduced back into the system with velocity components sampled according to a Maxwellian distribution centered on the wall velocity and random directions. Several models have been proposed to compute the equivalent force between DPD particles and solid walls[18] or the effective dissipative and random forces have been obtained analytically from the continuum limit of the interaction between fluid particles and wall particles.[17] These models usually must be combined with various reflection mechanisms to prevent liquid particles from penetrating through the walls.

Implementations of boundary conditions that involve collisions with sharp interfaces require algorithms that accurately locate interfaces and determine where and when particles reach them. This is straightforward for geometrically simple boundaries, but it is more challenging for complex stationary or moving boundaries that cannot be described by simple equations. The phase field approach[24-26] provides an accurate way to represent interfaces. It is based on the concept of a diffuse interface, can be defined in terms of a phase field, $\phi(\mathbf{x},t)$, that changes smoothly from one phase to the other over an interface zone with a non-zero width, *w*. In numerical applications, the parameters in the phase field equations are selected to ensure that the width of the interface is several times the size of the grid cell, which is used to define the phase field, to achieve a reasonable compromise between accuracy and efficiency. Beginning with applications to the solidifications of pure melts,[27-29] the phase field approach has been used to simulate a variety of interface dynamics phenomena (moving boundary problems) including solidification coupled with melt convection[30, 31], two-phase Navier-Stokes flow,[32] solute precipitation and/or dissolution,[33] diffuse-interface smoothed particle hydrodynamics (SPH) model for multiphase flow,[34] and grain growth. In most applications, the phase field equations are used to circumvent the difficulty of explicitly tracking sharp moving interfaces. In this paper, we use DPD simulations to show how a phase field method can be used in particle simulations to locate interfaces with subgrid scale resolution and determine when particles contact them.

## II. DPD AND PHASE-FIELD BASICS

### A. Dissipative Particle Dynamics

The DPD equations of motion are:

$$d\mathbf{r}_i / dt = \mathbf{v}_i, \tag{1}$$

$$m_i d\mathbf{v}_i / dt = \mathbf{f}_i = \sum_{j \neq i} \mathbf{f}_{ij}, \tag{2}$$

where $t$ is time, $\mathbf{f}_i$ is the force acting on particle $i$ and $m_i$, $\mathbf{r}_i$ and $\mathbf{v}_i$ are its mass, position vector and velocity vector. Here, $\mathbf{f}_{ij}$ is the force acting on particle $i$ due to its interaction with particle $j$. The symmetry, $\mathbf{f}_{ij} = -\mathbf{f}_{ji}$, between the particle-particle interactions ensures rigorous momentum conservation. Like the conservative force in MD, the DPD particle-particle interactions usually have a finite cutoff distance, $r_c$, and the summation runs only over all other particles, $j$, within the interaction cutoff, $r_c$. The particle-particle interactions, $\mathbf{f}_{ij}$, consist of three parts,

$$\mathbf{f}_{ij} = \mathbf{f}_{ij}^C + \mathbf{f}_{ij}^D + \mathbf{f}_{ij}^R, \tag{3}$$

where the superscripts *C*, *D* and *R* indicate the conservative, dissipative and randomly fluctuating forces.

The conservative force, $\mathbf{f}_{ij}^{C}$, can be written as,

$$\mathbf{f}_{ij}^{C} = a_{ij} w_{C}(r_{ij}) \hat{\mathbf{r}}_{ij}, \tag{4}$$

where, $a_{ij}$ is the interaction magnitude and $\mathbf{r}_{ij} = \mathbf{r}_i - \mathbf{r}_j$. $r_{ij} = |\mathbf{r}_{ij}|$ is the magnitude of $\mathbf{r}_{ij}$ and $\hat{\mathbf{r}}_{ij} = \mathbf{r}_{ij}/r_{ij}$. Since the cutoff distance $r_c$ is the only microscopic length in the system, it is used as the unit of length. In the standard DPD models, the weighting function $w_C$ has the form,

$$w_C(r) = \begin{cases} 1-r, & r < 1.0 \\ 0, & r \geq 1.0 \end{cases}. \tag{5}$$

The soft conservative force has a finite value with a maximum value of $a_{ij}$. The dissipative force $\mathbf{f}_{ij}^{D}$ represents the viscous forces between DPD particles, and it is assumed to depend only on the relative positions $\mathbf{r}_{ij}$ and relative velocities $\mathbf{v}_{ij} = \mathbf{v}_i - \mathbf{v}_j$ between particle *i* and *j*.

$$\mathbf{f}_{ij}^{D} = -\gamma w_D(r_{ij})(\hat{\mathbf{r}}_{ij} \cdot \mathbf{v}_{ij}) \hat{\mathbf{r}}_{ij}, \tag{6}$$

where the parameter $\gamma$ is a viscosity coefficient and $w_D$ is the weighting function for the dissipative forces. The random force component, $\mathbf{f}_{ij}^{R}$, represents the effects of thermal fluctuations, and it is usually written as,

$$\mathbf{f}_{ij}^{R} = \sigma w_R(r_{ij}) \xi_{ij} \hat{\mathbf{r}}_{ij}, \tag{7}$$

where $\sigma$ is a coefficient and $w_r$ is also an r-dependent weighting function. The randomly fluctuating variable, $\xi_{ij}$, in Eq. (7) is independent for each particle pair, $i$ and $j$, and it has a Gaussian distribution with $\langle \xi_{ij}(t) \rangle = 0$ and $\langle \xi_{ij}(t) \xi_{mn}(t') \rangle = (\delta_{im}\delta_{jn} + \delta_{in}\delta_{jm}) \delta(t-t')$. All three force components act along the line or centers between particles $i$ and $j$.

The coefficients and weighting functions of the dissipative and random forces are related through[12]

$$\gamma = \sigma^2/2k_B T, \tag{8}$$

where $k_B$ is the Boltzmann constant and

$$w_D(r) = \left[w_R(r)\right]^2, \tag{9}$$

in order to recover the correct thermodynamic equilibrium statistic at the prescribed temperature, $T$. In most DPD simulations, and in this work, the dissipative and random weighting functions

$$w_D(r) = \left[w_R(r)\right]^2 = \left[w_C(r)\right]^2 = \begin{cases} (1-r)^2, & r < 1.0 \\ 0, & r \geq 1.0 \end{cases}, \tag{10}$$

are used. The combination of dissipative and fluctuation forces are related through the fluctuation-dissipation theorem and act as a thermostat to maintain the temperature of the system, which can be measured through the average kinetic energy of the DPD particles. Therefore, DPD can be regarded as coarse grained thermostatted molecular dynamics. The modified velocity-Verlet algorithm used by Groot and Warren[35] was employed in this work.

**B. Phase-field representation of the liquid-solid interface**

In level set interface capturing,[36] the zero level set contour ($\phi = 0$) is used to implicitly represent the position of the sharp interface on a fixed grid, and the zero level contour of the phase field can be used in the same way if the phase field varies from -1 in phase 1 to +1 in phase 2 across the interface. In principle, both the phase-field approach and level set method can be used to represent any arbitrary interface through the variable $\phi(\mathbf{x},t)$. In the work presented in this paper, we used a phase-field function to illustrate how this approach can be used to implement solid-fluid boundary conditions.

The phase field model was originally developed as a theoretical approach to model and simulate multiphase materials, and it is based on the idea that the free energy of a two phase fluid can be described by a free energy density functional of the form

$$\mathcal{F}(\phi(\mathbf{x})) = F(\phi(\mathbf{x})) + \varepsilon^2/2 \, (\nabla \phi(\mathbf{x}))^2, \tag{11}$$

where $F(\phi)$ is the free energy density (free energy per unit volume) of a homogeneous systems characterized by the phase field, $\phi$, and the term $\varepsilon^2/2 \, (\nabla \phi)^2$ represents the contribution of density or composition gradients. If the homogeneous free energy has the form $F(\phi) = \phi^4/4 - \phi^2/2$, the relaxation of the phase field to the form that minimizes the total free energy ($\phi \approx 1$ in phase 1, $\phi \approx -1$ in phase 2 and $\phi$ varies continuously from –(1-δ) to 1-δ, with δ<<1, across a planar interface between the two phases in the direction perpendicular to the interface) is described by the equation

$$\tau \frac{\partial \phi}{\partial t} = \varepsilon^2 \frac{\partial^2 \phi}{\partial x^2} + \phi\left(1 - \phi^2\right), \tag{12}$$

where $\tau$ is a positive characteristic time constant and the coefficient $\varepsilon$ is closely related to the interface width. The dimensionless form of Eq. (12),

$$\frac{\partial \phi}{\partial t'} = \frac{\partial^2 \phi}{\partial x'^2} + \phi(1-\phi^2), \tag{13}$$

is obtained by introducing $t' = t/\tau$ and $x' = x/\varepsilon$, and the steady state one-dimensional stationary solution ($\partial \phi/\partial t' = 0$) to Eqs. (12) and (13) is,

$$\phi = \tanh(x'/\sqrt{2}) = \tanh(x/\sqrt{2}\,\varepsilon). \tag{14}$$

Starting from an initial step function, where $\phi = -1$ and $1$ represent the solid and liquid regions, the steady state planar interface profile shown in Fig 1 is obtained by integrating Eq. (13). If the interface is defined as the region in which $-0.9 < \phi < 0.9$, the interface width is $w \approx 3\sqrt{2}\varepsilon$, and the corresponding sharp interface is the $\phi = 0$ level set. In two and three dimensions with a curved interfaces, the governing equation for $\phi$, similar to Eq. (13), can be written as,

$$\frac{\partial \phi}{\partial t'} = \nabla'^2 \phi + \phi(1-\phi^2) - \kappa|\nabla'\phi|, \tag{15}$$

where $\nabla'^2 = \varepsilon^2 \nabla^2$ is the dimensionless Laplacian operator. The last term on the right-hand side of Eq. (15) counteracts the curvature driven interface motion[37]. The interface curvature can be calculated from $\kappa = \nabla' \cdot (\nabla'\phi/|\nabla'\phi|)$. Starting with an initial step function, the steady state phase field is established by solving Eq. (15) on a finite difference grid. The pseudo time $t'$ is used to evolve the phase field $\phi$ to steady state for any arbitrary solid-liquid interface. Figure 2 shows the interface (thick line), where $-0.9 < \phi < 0.9$, for a complex fracture geometry consisting of a self affine fractal and a replica that was translated both horizontally and vertically without rotation, after the phase field was relaxed by solving Eq. (15).

**C. Combination of DPD and phase-field interface representation**

To apply phase field interface representation to DPD simulations, a dense DPD particle fluid was initially generated and equilibrated in the entire computational domain at the selected temperature (Eq. (8)). Next, the value of the phase field, $\phi_i = \phi(\mathbf{r}_i)$, and the corresponding shortest distance from the interface,

$$d_i = \sqrt{2}\varepsilon \tanh(\phi_i) \tag{16}$$

was calculated for each particle, $i$, from linear interpolation of phase field $\phi$ on the underlying finite difference grid. Figure 3a shows the phase field values, $\phi_i$, for the DPD particles with the interface geometry shown in Fig. 2. The DPD particles representing the solid region are shown in dark gray and those representing the liquid are shown in light gray. If all of the particles with $\phi_i < \phi_i^{\min}$ are removed and the particles with $\phi_i^{\min} \leq \phi_i \leq 0$ are "frozen", the dynamics of the remaining particles with $\phi_i > 0$, the fluid particles, can be used to simulate fluid flow with solid boundaries. Figure 3b shows the resulting particle configuration with $\phi_i^{\min} = -0.9998$ (the fluid particles and solid particles in a thin boundary layer with a thickness of $1.08\varepsilon$).

Figure 4 shows how bounce-back (bb) reflection is used to impose no slip boundary conditions, and this serves as an illustration of how the phase-field interface representation can be used to implement collisions at interfaces with complex geometries. At any time, $t$, during the simulation, the position of particle $i$ is denoted by $\mathbf{r}_i^t$ and the particle phase-field variable, $\phi_i^t$, can be obtained by linear interpolation of $\phi$ from the underlying grid. The new position, $\mathbf{r}_i^{t+\Delta t}$, of particle $i$ after one time step of length $\Delta t$ can be obtained via time integration and thus the new particle phase field measure, $\phi_i^{t+\Delta t}$, can be updated at the new position, $\mathbf{r}_i^{t+\Delta t}$. If $\phi_i^{t+\Delta t} < 0$, the new particle position is inside the solid region and a bounce-back reflection is implemented to prevent particle $i$ from penetrating into the solid region. As a result of bounce-back reflection, the new velocity of particle $i$ is simply the reverse of the velocity before the bounce-back "collision", $\mathbf{v}_i^{t+\Delta t} = -\mathbf{v}_i^t$, for a zero wall velocity $\mathbf{v}_{wall} = 0$. For a nonzero wall velocity, in Couette flow for example, the particle velocity after bounce back reflection is given by $\mathbf{v}_i^{t+\Delta t} = -\mathbf{v}_i^t + 2\mathbf{v}_{wall}$. The distances $AC$ and $CD$ in Fig. 4 can be calculated using the particle phase-field information from Eq. (16) as $AC = \sqrt{2}\varepsilon \cdot a\tanh(\phi_i^t)$ and $CD = \sqrt{2}\varepsilon \cdot a\tanh(-\phi_i^{t+\Delta t})$. The new position after bounce-back, $\mathbf{r}_{i,bb}^{t+\Delta t}$, can easily be calculated geometrically from

$$\mathbf{r}_{i,bb}^{t+\Delta t} = \mathbf{r}_i^{t+\Delta t} - 2\mathbf{v}_i^t \Delta t \left[ a\tanh\left(-\phi_i^{p,t+\Delta t}\right) \right] / \left[ a\tanh\left(-\phi_i^{p,t+\Delta t}\right) + a\tanh\left(\phi_i^{p,t}\right) \right]. \quad (17)$$

The new particle phase-field variable $\phi_{i,bb}^{t+\Delta t}$ can be updated at the new particle position after bounce-back, $\mathbf{r}_{i,bb}^{t+\Delta t}$, and usually $\phi_{i,bb}^{t+\Delta t} \approx -\phi_i^{t+\Delta t}$, which is positive indicating that the new position after bounce-back is in the liquid region. For moving solid walls, the velocity $\mathbf{v}_i^t$ must be replaced by the relative velocity $\left(\mathbf{v}_i^t - \mathbf{v}_{wall}\right)$ in Eq. (17). The entire procedure must be implemented for all DPD particles that are close enough to the interface to reach it in a single time step (those with $\phi_i$ close to 0). For DPD particles in the bulk fluid with $\phi_i \square 0$, it is not necessary to repeat this procedure at every time step as long as $\Delta t$ is small enough, and this reduces the computational cost.

## III. NUMERICAL EXAMPLES

## A. Poiseuille flow

To test the implementation of phase-field enabled bounce-back boundary conditions, 2D DPD simulations of Poiseuille flow in a narrow channel with a width of 10 were performed. Periodic boundary conditions were imposed along the flow direction (the $x$ direction). The system contained 1440 DPD particles randomly distributed in the simulation domain. Particles with $\phi_i < 0$ were designated as solid particles and particles with $\phi_i > 0$ were designated fluid particles. Figure 5 shows a snapshot of the computational domain and the phase field profile across the channel. An external body force equal to $g = 0.02$ (in DPD units) was imposed on each fluid particle to initiate and sustain the flow. The densities of the fluid and solid were 4, and the prescribed temperature was $k_B T = 1.0$. The dissipative and random coefficients were chose to be $\gamma = 4.5$ and $\sigma = 3.0$ to satisfy the fluctuation-dissipation theorem constraint. The conservative force parameter between fluid-fluid particles was set to $a_{ff} = 18.75$ to match the compressibility of water,[35] and the fluid-wall interaction was $a_{fs} = 18.75$ - the same as the fluid-fluid interaction. The flow domain was divided into 100 bins across the channel, and data was collected and averaged for each bin after the system reached steady state. Under laminar, low Reynolds number, flow conditions, the velocity profile across the channel is given by:

$$v_x(y) = \frac{\rho g}{2\mu}(a^2 - y^2) , \qquad (18)$$

where $\mu$ is the shear viscosity and $a$ is the half width of the channel. Figure 6 presents the DPD density, temperature and stream velocity results, as well as the velocity profile given by Eq. (18). The density and temperature are essentially constant across the channel, as they should be, and the no-slip boundary conditions are well satisfied. Deviations from the nominal temperature of 1.0 indicate integration errors, and these deviations are small in this simulation.

The shear viscosity $\mu$ for the DPD fluid can be estimated from either the maximum or the average stream velocity,

$$\mu = \frac{\rho g a^2}{2 v_x^{max}} \text{ or } \mu = \frac{\rho g a^2}{3 \bar{v}_x}, \tag{19}$$

where the average fluid velocity $\bar{v}_x$ can be calculated by averaging all fluid DPD particle velocities along the $x$ direction. The DPD fluid dynamic viscosity was estimated to be $\mu = 0.926$ and $\mu = 0.912$ from the two expressions given in Eq. (19), with a Reynolds number of $R_e = 31.3$. The same simulation was also run for $k_B T = 0.1$ with a conservative fluid-fluid particle interaction parameters of $a_{ff} = 1.875$ and $\gamma = 45.0$. The viscosity was estimated to be $\mu = 3.4072$ and $\mu = 3.4066$, and the Reynolds number of the flow was $R_e = 2.3$. This viscosity value was used for the simulations of flow through parallel cylinders in the next section.

**B. Flow through an array of parallel cylinders**

A second numerical example was implemented to test the accuracy of DPD simulations with a more complex geometry. Stokes flow through a square array of parallel cylinders was simulated using DPD with phase field implementation of the boundary reflections. Periodic boundary conditions were imposed along both the $x$ and $y$ directions. The system contained 3120 DPD particles randomly distributed in the simulation domain of size $L \times L$. The solid particles forms a circle of radius $r$ (Fig. 7a), and an external body force equal to $g = 0.005$ (DPD units) was imposed on each fluid particle along the $x$ direction to initiate and sustain the flow. The analytical solutions of the normalized mean fluid velocity along the $x$ direction $\overline{u}_x$ is,[38]

$$\overline{u}_x = \frac{4\pi \overline{v}_x \mu}{\rho g L^2} = \ln\sqrt{\frac{\pi}{\varepsilon}} + \varepsilon - 1.31 - 8.76\left(\frac{\varepsilon}{\pi}\right)^2 + 63.22\left(\frac{\varepsilon}{\pi}\right)^3, \tag{20}$$

where $\overline{v}_x$ is the mean velocity before normalization, and $\varepsilon = \pi r^2/L^2$ is the void fraction. Higher order terms (>3) in the original analytical solution in ref. [38] are neglected. Simulations were run for several void fractions, $\varepsilon$, at a DPD temperature of $k_B T = 0.1$. The fluid dynamic viscosity was estimated to be $\mu = 3.41$ from the Poiseuille flow simulations. The entire computational domain was divided into 25x25 small bins and the velocity was averaged for each bin. A typical flow field around a cylinder for a void fraction of $\varepsilon = 0.1$ is shown in Fig. 7(b). The mean fluid velocity $\overline{u}_x$ was computed after the system reached a steady state, and the results are compared to the analytical solution in Fig. 8. The DPD results are in good agreement with the analytical solution (Eq. (20)). Eq. (20) is not valid for large void fractions and a modified solution can be constructed by keeping the first term on the right-hand-side of Eq. (20) and varying the coefficients of the polynomial in ε to achieve the best agreement with the DPD simulation result over the entire range of void fractions. The modified equation is

$$\overline{u}_x = \frac{4\pi \overline{v}_x \mu}{g L^2} = \ln\sqrt{\frac{\pi}{\varepsilon}} + 1.7\varepsilon - 1.31 - 2.15\varepsilon^2 + 1.35\varepsilon^3. \tag{21}$$

Eq. (21) incorporates the physics from the analytical solution, namely $\overline{u}_x \propto -\ln(\varepsilon)$ at small void fraction, $\varepsilon$, and it was extend to the entire range of void fractions investigated. This empirical relationship is also plotted in Fig. 8.

**C. Unsaturated flow through porous media**

An attractive feature of the phase field method is the relatively more accurate representation of the boundary and simpler implementation of the boundary reflections for particle models through the phase-field variable $\phi$. This makes it very useful for simulating flows in complex confined geometries using particle methods, for example, flow through porous media using DPD or SPH (smoothed particle hydrodynamics), where the liquid-solid boundaries are extremely complicated and an efficient and accurate scheme for imposing no-slip boundary conditions is essential.

The last numerical example is a DPD simulation of unsaturated flow through a two-dimensional fractured porous medium with phase-field implementation of the boundary reflections. The porous medium was represented by a set of circular disks of different radii, which were sampled from a prescribed size distribution and randomly positioned within the model domain, avoiding overlap with previously inserted discs. In order to simulate a fractured porous medium discs were not inserted if their centers lay inside any other discs. Periodic boundary conditions were imposed along both the $x$ and $y$ directions. A DPD simulation was first run in the entire computational domain and solid particles that are not within a preselected small distance from the nominal solid-fluid interface were removed. In this study, solid particles at locations where $-0.9998 < \phi_i < 0$ form the fluid-solid boundary (in blue colors) and the rest of the solid particles with $\phi_i \leq -0.9998$ were removed to save computational expense (Fig. 9). Liquid initially filled the left part of the computational domain and an external body force equal to $g = 0.02$ (DPD units) acting along the $x$ direction was imposed on each fluid particle to initiate and sustain the flow.

The standard DPD model with purely repulsive interactions between DPD particles has been used extensively to simulate the behavior of multiphase fluids in confined systems. A more realistic DPD model with a combination of long range attractive and short range repulsive interactions can be used to simulate the behavior of unsaturated flow in various confined geometries,[39] and this approach was used in this simulation. For example, the interaction energy between two DPD particles, $i$ and $j$, can be represented by,

$$E(r_{ij}) = S_{ij}^r W(r_{ij}, r_{0,ij}^r) - S_{ij}^a W(r_{ij}, r_{0,ij}^a), \tag{22}$$

where $W(r, r_0)$ is a smooth function with a support scale of $r_0$, similar to the smoothing function used in smoothed particle hydrodynamics. Here $S_{ij}^r$ and $S_{ij}^a$ are the strengths of the repulsive and attractive interactions between particle $i$ and $j$, and $r_{0,ij}^r$ and $r_{0,ij}^a$ are the corresponding cutoff range for repulsive and attractive interactions. The form of the function $W(r, r_0)$ used in this work was

$$W(v, h) = \frac{\alpha_D}{h^D} \begin{cases} (1 - 3/2\,v^2 + 3/4\,v^3) & 0 \leq v < 1 \\ 1/4(2-v)^3 & 1 \leq v < 2 \\ 0 & otherwise \end{cases}, \tag{23}$$

(a widely used smoothing function in SPH), where $h = r_0/2$ and $v = r/h$. $D$ is the spatial dimensionality, and $\alpha_D$ is a constant that assures the proper normalization of the smoothing function ($\alpha_D = 2/3, 10/7\pi, 1/\pi$ for D = 1, 2 and 3). Corresponding conservative interaction forces can be obtained by taking the derivatives of Eq. (22),

$$f_i^c = \sum_{j \neq i} \left[ S_{ij}^r W'(r_{ij}, r_{0,ij}^r) - S_{ij}^a W'(r_{ij}, r_{0,ij}^a) \right] \hat{\mathbf{r}}_{ij}, \tag{24}$$

where $W'(r, r_0) = dW(r, r_0)/dr$.

In the simulation, a temperature of $k_B T = 0.5$ and a dissipation (friction) coefficient of $\gamma = 4.5$ were used. The parameters for the fluid-fluid particle interaction were $S_{ff}^r = 37.5$, $S_{ff}^a = 18.75$, $r_{0,ff}^r = 0.73$, and $r_{0,ff}^a = 1.0$. The strength of the fluid-solid interaction was twice the strength of the fluid-fluid interaction and consequently the fluid wets the solid walls. Figure 9 shows snapshots of a DPD simulation of gravity driven unsaturated flow across the porous medium. It is difficult to make direct comparisons with a grid-based simulation using Navier-Stokes solvers without extensive "calibration" simulations to determine the viscosity and surface tension of the DPD fluid. Matching the wetting behavior between DPD and continuum models is even more challenging. Mesoscopic DPD simulations capture the full complexity of the fluid-fluid-solid contact line (contact point in two-dimensional simulations) including fluctuating local contact angles that depend on the velocity of the fluid relative to the solid. Most continuum models assume a fixed contact angle or a very small number of contact angles (typically an advancing contact angle and a smaller receding contact angle). Because of the soft nature of particle-particle interaction, a few particles with high kinetic energy can penetrate into the solid region, and a boundary reflection algorithm must be implemented at the fluid-solid interface to prevent penetration. The complicated geometry can make it very difficult to determine when a fluid particle has reached the boundary. The phase field method provides a very convenient, robust, and accurate way of implementing boundary reflections at geometrically complex interfaces.

## V. CONCLUSIONS

A phase field approach provides a simple, robust and effective way of implementing collision boundary conditions in particle model simulations of fluids in confined systems with complex boundary geometries. The method was illustrated for DPD simulations with no slip boundaries implemented using bounce-back reflection, and validated by simulations of Poiseuille flow and Stokes flow through a square array of parallel solid cylinders. An application of the method to a system with a more complicated geometry, unsaturated flow through a porous medium, was also presented. The method was also successfully applied in earlier investigations of multiphase unsaturated flow through straight channel, complex fractures, and fracture junctions.[40]

**ACKNOWLEDGMENTS**

This work was supported by the U.S. Department of Energy, Office of Science Scientific Discovery through Advanced Computing Program. The Idaho National Laboratory is operated for the U.S. Department of Energy by the Battelle Energy Alliance under Contract DE-AC07-05ID14517.

**References**

[1] A. G. Schlijper, P. J. Hoogerbrugge and C. W. Manke, J. Rheol. **39**, 567-579 (1995).

[2] Y. Kong, C. W. Manke, W. G. Madden and A. G. Schlijper, J. Chem. Phys. **107**, 592-602 (1997).

[3] E. S. Boek, P. V. Coveney and H. N. W. Lekkerkerker, J. Phys.: Condens. Matter **8**, 9509-9512 (1996).

[4] E. S. Boek, P. V. Coveney, H. N. W. Lekkerkerker and P. vanderSchoot, Phys. Rev. E **55**, 3124-3133 (1997).

[5] P. V. Coveney and K. E. Novik, Phys. Rev. E **54**, 5134-5141 (1996).

[6] P. G. Degennes, Rev. Mod. Phys. **57**, 827-863 (1985).

[7] P. Meakin and A. M. Tartakovsky, Rev. Geophys.

[8] B. P. Binks, Curr. Opin. Colloid Interface Sci. **7**, 21-41 (2002).

[9] W. Kang and U. Landman, Phys. Rev. Lett. **98**, 064504 (2007).

[10] P. J. Hoogerbrugge and J. M. V. A. Koelman, Europhys. Lett. **19**, 155-160 (1992).

[11] R. Kubo, Rep. Prog. Phys. **29**, 255-282 (1966).

[12] P. Espanol and P. Warren, Europhys. Lett. **30**, 191-196 (1995).

[13] I. V. Pivkin and G. E. Karniadakis, J. Chem. Phys. **124**, 184101 (2006).

[14] J. G. Kirkwood, J. Chem. Phys. **7**, 919 (1939).

[15] B. J. Alder and T. E. Wainwright, Phys. Rev. **127**, 359 (1962).

[16] W. Dzwinel and D. A. Yuen, Int. J. Mod. Phys. C **11**, 1-25 (2000).

[17] M. Revenga, I. Zuniga and P. Espanol, Comput. Phys. Commun. **122**, 309-311 (1999).

[18] I. V. Pivkin and G. E. Karniadakis, Journal of Computational Physics **207**, 114-128 (2005).

[19]S. M. Willemsen, H. C. J. Hoefsloot and P. D. Iedema, Int. J. Mod. Phys. C **11**, 881-890 (2000).

[20]M. Revenga, I. Zuniga, P. Espanol and I. Pagonabarraga, Int. J. Mod. Phys. C **9**, 1319-1328 (1998).

[21]W. Gaede, Ann. Phys.-Berlin **41**, 337-380 (1913).

[22]P. S. Epstein, Phys. Rev. **23**, 710-733 (1924).

[23]R. A. Millikan, Phys. Rev. **22**, 1-23 (1923).

[24]J. D. van der Waals, Verh.-K. Ned. Akad.Wet., Afd. Natuurkd., Eerste Reeks **1**, 8 (1893).

[25]V. L. a. L. Ginzburg, L. D., Zhurnal Eksperimentalnoy i Teoreticheskoy Fizicheskoi **20**, 1064-1082 (1950).

[26]J. W. Cahn and J. E. Hilliard, J. Chem. Phys. **28**, 258-267 (1958).

[27]A. Karma and W. J. Rappel, Phys. Rev. E **53**, R3017-R3020 (1996).

[28]J. B. Collins and H. Levine, Phys. Rev. B **31**, 6119-6122 (1985).

[29]J. S. Langer, *Directions in Condensed Matter* (World Scientific, Philadelphia, 1986).

[30]D. M. Anderson, G. B. McFadden and A. A. Wheeler, Physica D **135**, 175-194 (2000).

[31]C. Beckermann, H. J. Diepers, I. Steinbach, A. Karma and X. Tong, J. Comput. Phys. **154**, 468-496 (1999).

[32]D. Jacqmin, J. Comput. Phys. **155**, 96-127 (1999).

[33]Z. Xu and P. Meakin, J. Chem. Phys. **129**, 014705 (2008).

[34]Z. Xu, P. Meakin and A. M. Tartakovsky, Phys. Rev. E **79**, 036702 (2009).

[35]R. D. Groot and P. B. Warren, J. Chem. Phys. **107**, 4423-4435 (1997).

[36]J. A. Sethian and P. Smereka, Annu. Rev. Fluid Mech. **35**, 341-372 (2003).

[37]R. Folch, J. Casademunt, A. Hernandez-Machado and L. Ramirez-Piscina, Phys. Rev. E **60**, 1724-1733 (1999).

[38]J. E. Drummond and M. I. Tahir, Int. J. Multiph. Flow **10**, 515-540 (1984).

[39]M. B. Liu, P. Meakin and H. Huang, Physics of Fluids **18**, 017101 (2006).

[40]P. Meakin and Z. Xu, Prog. Comput. Fluid Dyn. (2009).

FIG. 1. Phase-field profile across an interface. The phase field, $\phi$, varies continuously across the interface from -1 (solid region) to +1 (liquid region). The interface width, $w \approx 3\sqrt{2}\varepsilon$, corresponds to the width of the region in which $-0.9 < \phi < 0.9$.

FIG. 2. A phase-field interface representation of a complex fracture. The thick solid lines represent the liquid-solid interfaces.

FIG. 3. a) DPD particle configuration based on the phase-field representation in Fig. 2. The particles are shaded according to the values of the phase-field, $\phi^p$ at the particles. b) A thin layer of solid particles is "frozen" near the liquid-solid interface and all other solid particles, those far from the interface, have been removed.

FIG. 4. A cartoon of the implementation of DPD bounce-back reflection based on a phase-field description of the liquid-solid interface.

FIG. 5. DPD Poiseuille flow. Left: a snapshot of a DPD Poiseuille flow simulation, with particles colored according to the phase-field variable, $\phi^p$. Right: $\phi^p$ profile across the entire channel.

FIG. 6. DPD simulation results for Poiseuille flow compared with the Navier-Stokes solution. The solid walls are simulated by freezing DPD particles in the solid region in combination with bounce-back reflection as described in the text.

FIG. 7. DPD Stokes flow through a square array of parallel cylinders. Left: a snapshot of a typical particle configuration with the DPD particles shaded according to the phase-field variable, $\phi^p$. Right: a typical velocity vector field from a DPD simulation.

FIG. 8. DPD results and the analytical solution (Eq. (20)) for the dependence of the normalized mean velocity, $\bar{u}_x$, on the void fraction, $\varepsilon$. The DPD results are in good agreement with the analytical solution for small void fraction. The analytical solution is not valid for large void fraction. The empirical fitted solution (Eq. (21)) is presented as thick dash-dot line.

FIG. 9. Snapshots of DPD gravity driven unsaturated flow through porous media. Gravity acts from left to right. a) t = 0; b) t = 100; c) t = 200; d) t = 300.

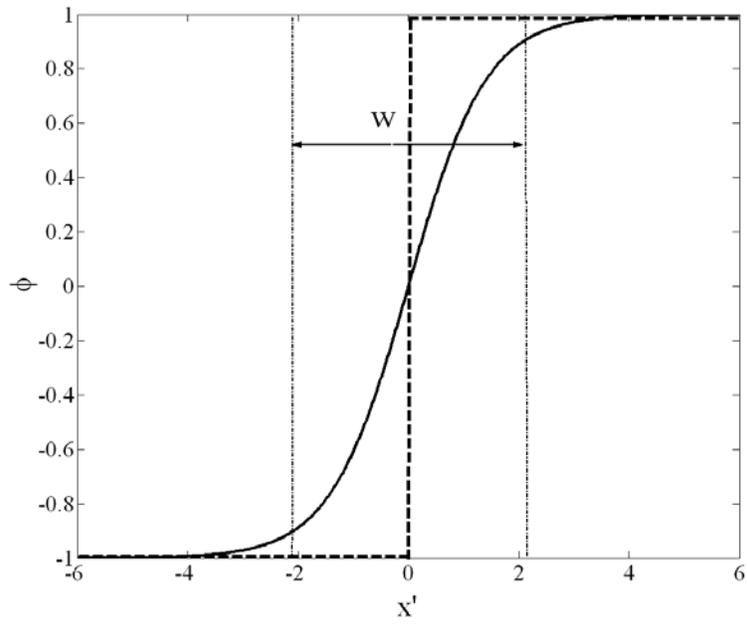

FIG. 1.

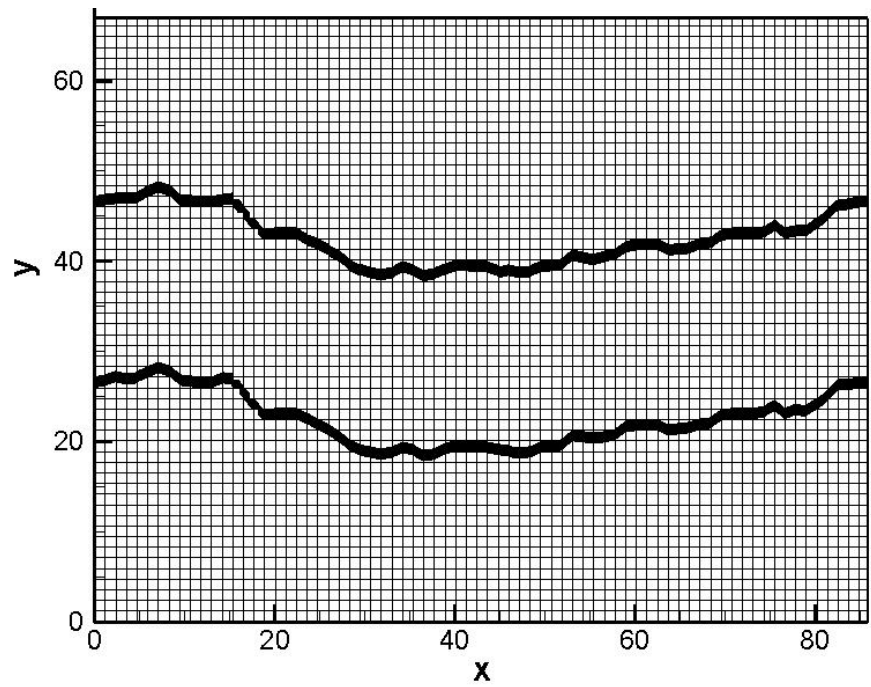

FIG. 2.

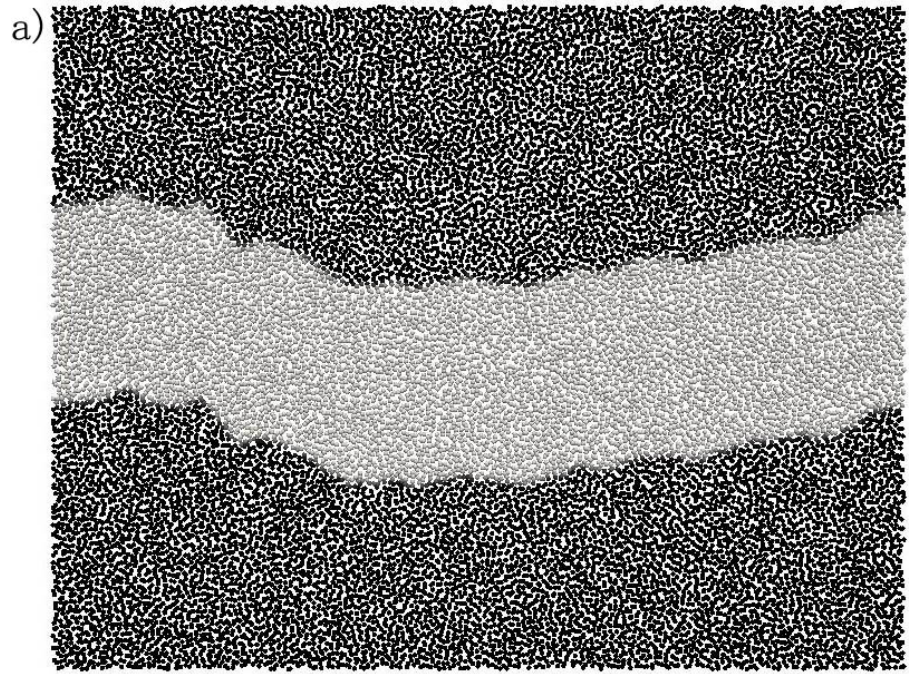

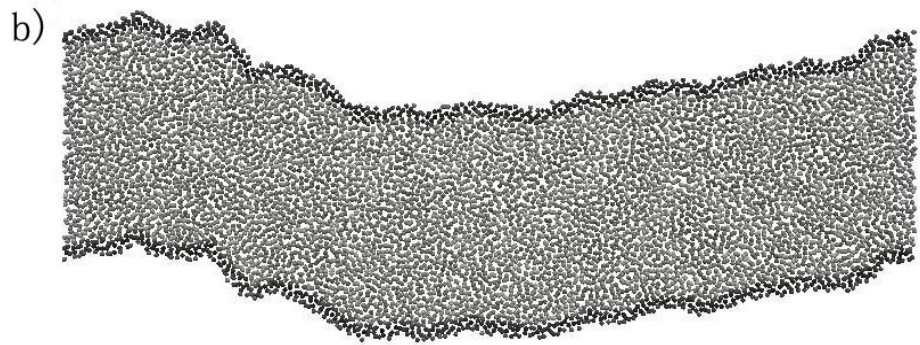

FIG. 3.

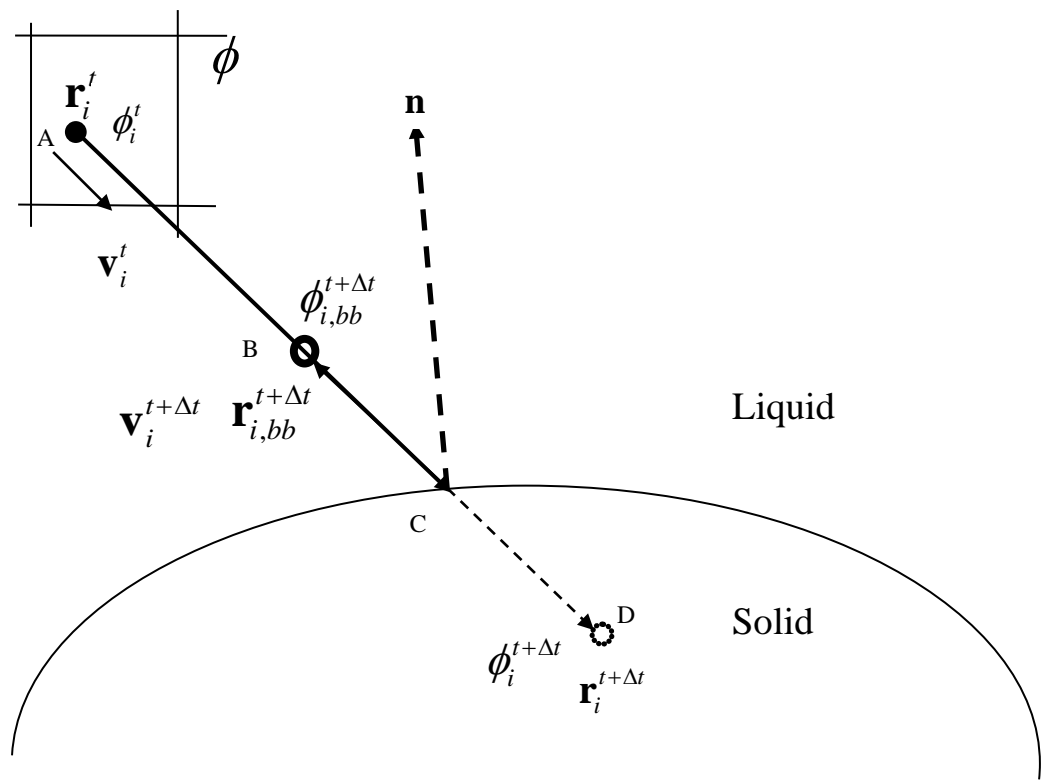

FIG. 4.

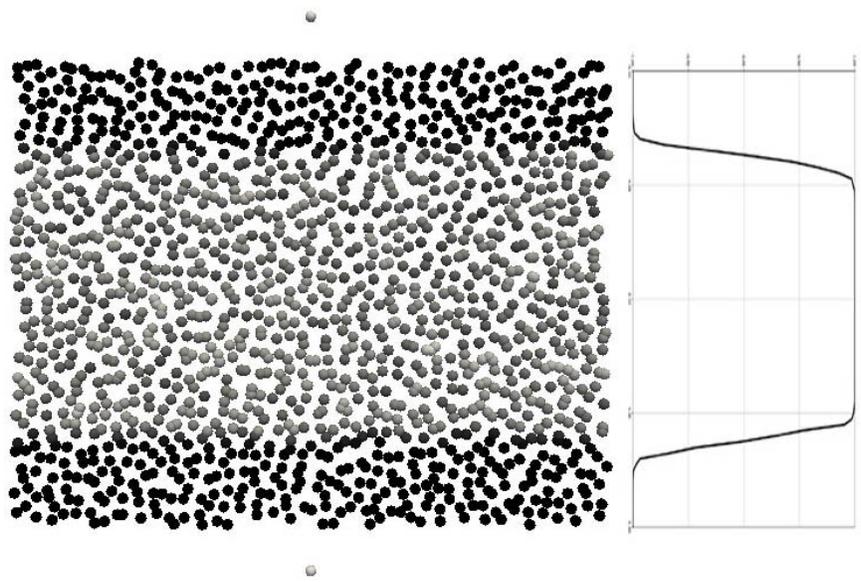

FIG. 5

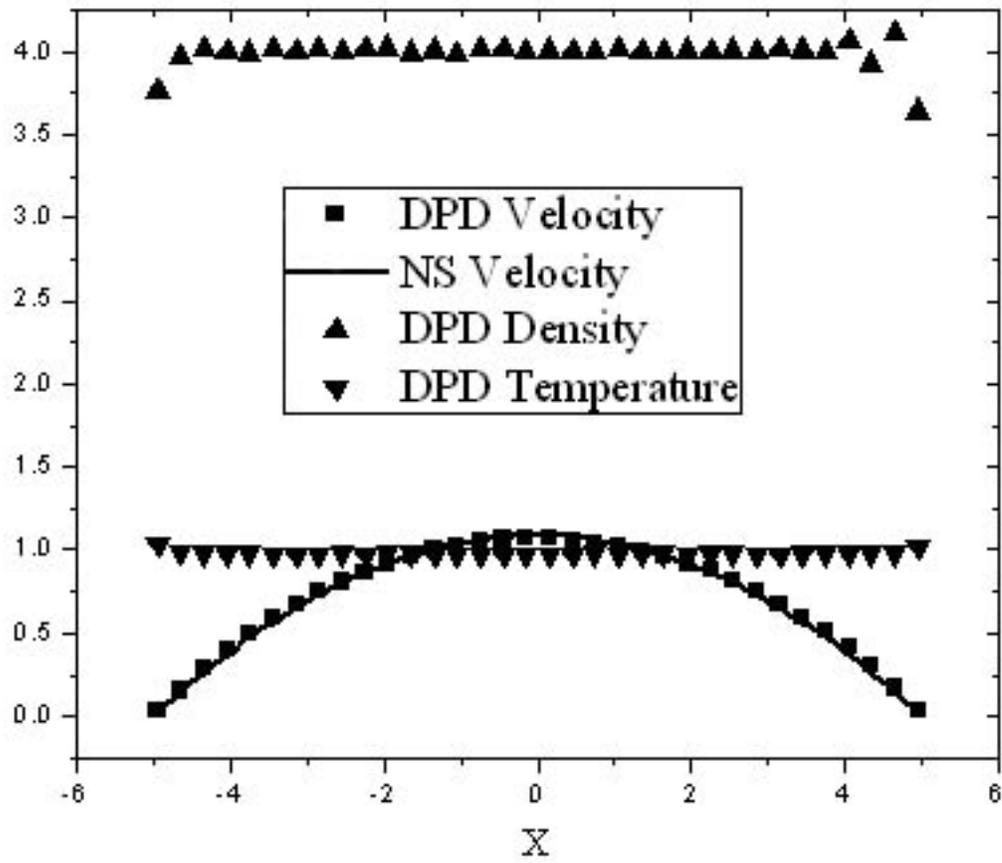

FIG. 6

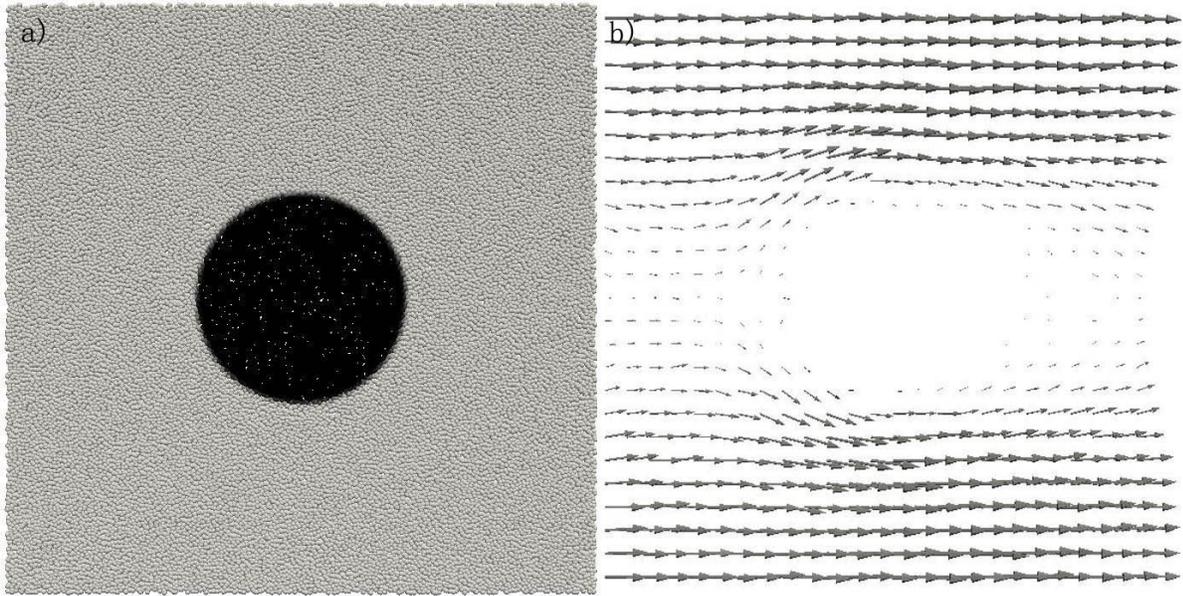

FIG. 7

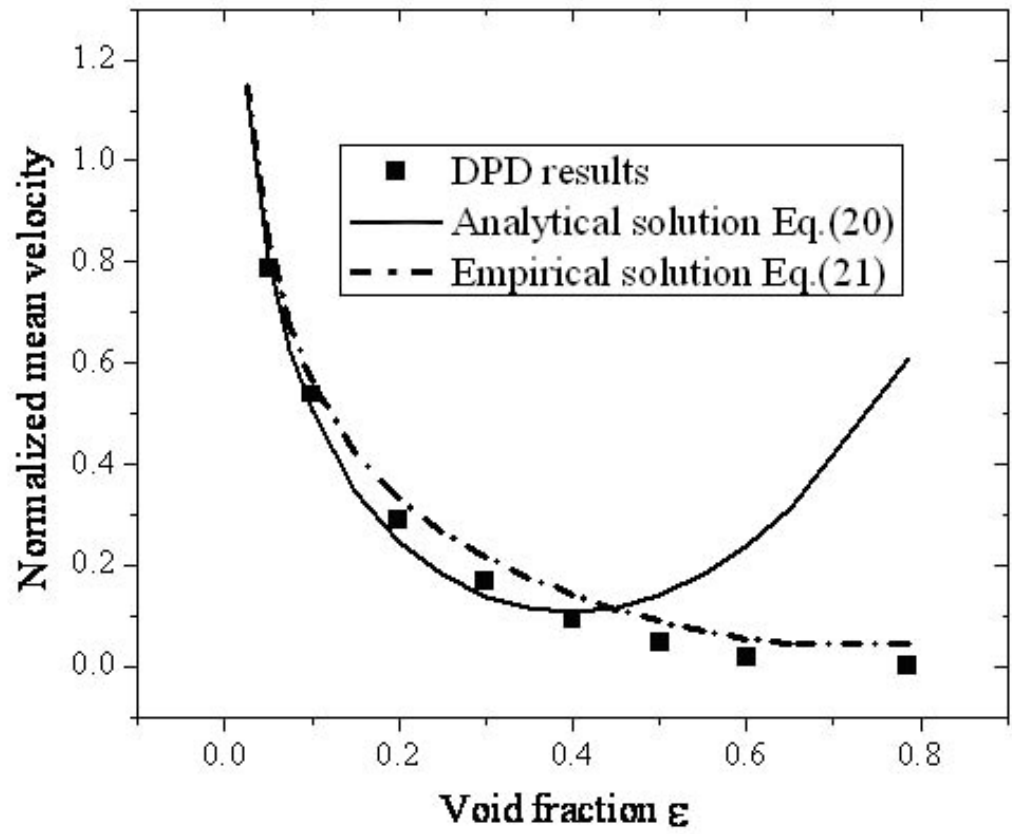

FIG. 8

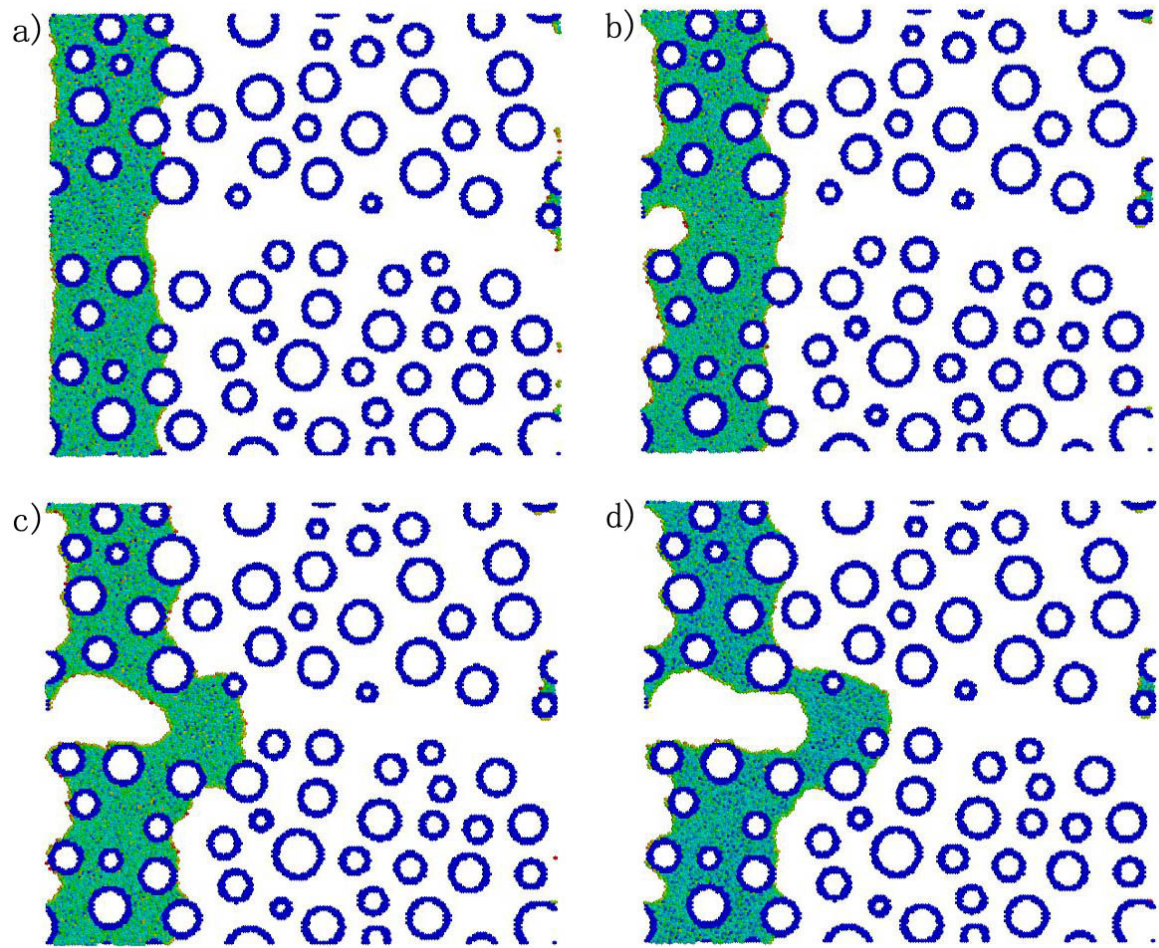

FIG. 9